\documentclass[preprint,aps,prl,groupedaddress,floatfix,showpacs,showkeys]{revtex4}

\usepackage{graphicx,amsmath,amssymb,dcolumn}
\usepackage{bm}

\newcommand{\bnmr}{$\beta$-NMR }

\preprint{APS/nucl-ex}

\begin{document}

\title{Ground state magnetic dipole moment of $^{35}$K}

\author{
T.J.~Mertzimekis$^{1,}$\footnote{Corresponding author, email: mertzime@nscl.msu.edu},
P.F.~Mantica$^{1,2}$,
A.D.~Davies$^{1,3}$,
S.N.~Liddick$^{1,2}$,
B.E.~Tomlin$^{1,2}$
}
\affiliation{
$^{(1)}$ National Superconducting Cyclotron Laboratory, Michigan State
University, East Lansing, MI 48824
}
\affiliation{
$^{(2)}$ Department of Chemistry, Michigan State University, East Lansing, MI 48824
}
\affiliation{
$^{(3)}$ Department of Physics and Astronomy, Michigan State
University, East Lansing, MI 48824
}

\date{\today}

\begin{abstract}
\label{abstract}
The ground state magnetic moment of $^{35}$K has been measured using
the technique of nuclear magnetic resonance on $\beta$-emitting
nuclei. The short-lived $^{35}$K nuclei were produced following the
reaction of a $^{36}$Ar primary beam of energy 150~MeV/nucleon
incident on a Be target. The spin polarization of the $^{35}$K nuclei
produced at 2$^\circ$ relative to the normal primary beam axis was
confirmed. Together with the mirror nucleus $^{35}$S, the measurement
represents the heaviest $T=3/2$ mirror pair for which the spin
expectation value has been obtained. A linear behavior of $g_p$
vs. $g_n$ has been demonstrated for the $T=3/2$ known mirror moments
and the slope and intercept are consistent with the previous analysis
of $T=1/2$ mirror pairs.
 
\end{abstract}


\maketitle


\section{Introduction}
\label{intro}

The magnetic dipole moment $\mu$ provides important details on orbital
and spin contributions to nuclear state wave functions, mainly due to its
sensitivity to the individual proton and neutron contributions. The
extreme limits of the magnetic moments are represented by the
so-called Schmidt values and the vast majority of the known magnetic
moments fall within those extremes. One special aspect of magnetic
moments can be applied to understand gross spin properties of nuclei,
that is, the sum of the ground-state magnetic moments of mirror
nuclei. This sum represents the isoscalar part of the magnetic moment
multiplied by a factor of two, and it can be directly related to the
Pauli spin expectation value, $\displaystyle
\langle \sum_i \sigma^i_z\rangle$ or $\langle
\sigma\rangle$, as illustrated in the following
relation~\cite{Sugimoto73:spin}:
\begin{eqnarray}
\mu(T_z=+T) + \mu(T_z=-T)=&~\nonumber\\
 J + (\mu_\pi & + \mu_\nu -\frac{1}{2})~\langle{\displaystyle \sum_{i}} \sigma^i_z\rangle
\label{eq:spin}
\end{eqnarray}


In the above, $\mu_{\pi(\nu)}$ is the free proton (neutron) magnetic
moment in units of the nuclear magneton, $\mu_N$, $J$ is the total
angular momentum and $T$ represents the isospin. Eqn.~\ref{eq:spin}
is valid {\em only} if isospin is a good quantum number, providing a
means to test isospin symmetry breaking.

A limiting factor to test isospin symmetry is the difficulty to
measure magnetic moments far from stability. Mirror nuclei lie close
to the $N=Z$ line and become unstable above $A\sim 40$. As mass
increases, it is more difficult to overcome the challenges of the
available moment measurement techniques imposed by short lifetimes and
low production rates. The situation is especially pronounced for the
case of neutron-deficient nuclei with isospin $T_z=-3/2$. While mass
$A=43$ has been reached for $T=1/2$, magnetic moments for $T=3/2$
mirror nuclei are only known up to $A=17$~\cite{Geithner05:ne_g}. As
$Z$ increases, Coulomb repulsion effects within the nucleus become
more significant. At these higher masses, the increasing nuclear
charge might have a direct effect on isospin, potentially leading to
symmetry breaking.

Structural effects play an immediate role in determining the value of
the magnetic moment. The reverse also holds: a measurement of
the magnetic moment can reveal critical information about the
structure of the nucleus under investigation. Several endonuclear
effects, such as core polarization and meson current
exchange~\cite{Yamazaki70:meson,Arima73:core}, have been found to
alter the magnetic moment operator at a level which can be observed
via a magnetic moment measurement. Along this line, empirical
relationships have been established for either ground
state~\cite{Buck83:older,Buck01:newer} or excited
state~\cite{Mertzimekis03:a80} magnetic moments to overcome the lack
of a firm theoretical explanation.

The present work focuses on extending the known measurements of
$T=3/2$ nuclei. Recently, we have demonstrated that significant spin
polarization can be generated in single-nucleon pickup reactions at
intermediate energies~\cite{Groh03:k37}. Spin polarization was studied
in $^{37}$K isotopes created as products in single proton-pickup
reactions. The polarization was observed to be maximum near the peak
of the fragment momentum distribution reaching a relatively large
magnitude of $\sim 8.5\%$.

The polarization produced in the proton-pickup reaction may be a key
factor for extending magnetic moment measurements to heavier
$T_z=-3/2$ nuclei. Therefore we have employed a charge-pickup $^{36}$Ar($p$,$2n$)
reaction, where the produced $^{35}$K nuclei ($J^\pi=\frac{3}{2}^+$,
$t_{1/2}=190$~ms, $Q_{EC}=11881$~keV) are expected to exhibit
considerable polarization. The effect of neutron evaporation during
the reaction on the observed polarization due to two fewer neutrons
with respect to $^{37}$K is an open question.

A previous measurement of $\mu$($^{35}$K) was attempted at GSI by
Sch{\" a}fer {\em et al.}~\cite{Schaefer98:k35} via a
fast-fragmentation reaction. The experimental result suffered small
polarizations and low counting statistics and a $g$ factor, $g=0.24(2)$,
was extracted. Sch{\" a}fer {\em et al.} also
suggested that the systematics of $T=3/2$ nuclei did {\em not} follow
the linear behavior between the effective $g$ factors of the
proton-odd and neutron-odd nuclei of the mirror pair, $\gamma _p$ and
$\gamma _n$ respectively, as observed in the case of $T=1/2$ nuclei by
Buck and Perez~\cite{Buck83:older,Buck01:newer}. We report on the
asymmetry of $^{35}$K nuclei produced via a ($p$,$2n$) proton-pickup
reaction and a more precise value of the ground state magnetic moment
of $^{35}$K measured using the $\beta$-NMR technique. This new value
of $\mu$($^{35}$K), in combination with $^{35}$S mirror
data~\cite{Burke54:s35} is used to examine the Pauli spin expectation
value $\langle \sigma\rangle$ and the relation between $\gamma_p$ and
$\gamma_n$ for $T=3/2$ nuclei.
\begin{figure}
\begin{center}
\includegraphics[
width=8cm,
]{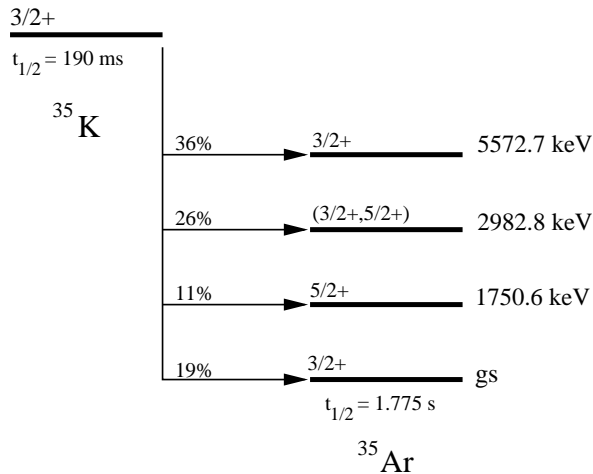}
\end{center}
\caption{Decay of $^{35}$K to $^{35}$Ar, where only the strongest
branchings are shown~\cite{ToI:98}. The energy levels are not in scale.}
\label{fig:k35decay}
\end{figure}


\section{Experimental Technique}

A primary beam of $^{36}$Ar was accelerated to 150 MeV/nucleon by the
coupled cyclotrons at the National Superconducting Cyclotron
Laboratory and impinged on a 564 mg/cm$^2$ Be target to create a
secondary $^{35}$K beam via the proton-pickup reaction
$^{36}$Ar($^{9}$Be,$^{10}$Li)$^{35}$K. Two dipole magnets located
upstream of the Be production target were used to steer the primary beam
to an angle of 2$^\circ$ with respect to the normal beam axis. The
$^{35}$K nuclei were separated from the reaction products using the
A1900 fragment separator~\cite{Morrisey03:A1900}. The desired $^{35}$K
isotopes were finally delivered to the $\beta$-NMR end station at
energies around 50~MeV/nucleon and a rate of $\sim$30~pps/pnA. The
main contaminant was $^{34}$Ar and the implantation ratio between
$^{35}$K and $^{34}$Ar was roughly 1:1.

The A1900 is able to deliver beams over a broad range of both angular
and momentum acceptance. The full momentum acceptance of the device is
5$\%$, while the measured angular acceptance is 60~mrad in the
horizontal direction and 40~mrad in the vertical direction. In the
present experiment, the full angular acceptance was selected. The
rigidity of the last two dipole magnets downstream in the fragment
separator were set to $B\rho_{3,4}=1.6910$~Tm. The particle
identification was achieved using standard energy loss and
time-of-flight measurements.

At the exit of the beam line, the secondary beam passed through a
Kapton window and traveled through air for 20 cm before being
implanted into a KBr single crystal located at the center of the \bnmr
apparatus. The \bnmr apparatus~\cite{Mantica97:bNMR} consisted of a
large dipole magnet with its poles perpendicular to the beam direction
and a distance of 10~cm between them. The magnet provides the required
Zeeman hyperfine splitting of the levels of the spin-polarized nuclei.
Two $\beta$ telescopes, each consisting of a thin
$\Delta E$ (4.4~cm $\times$ 4.4~cm $\times$ 0.3~cm) and a thick
$E$ (5.1~cm $\times$ 5.1~cm $\times$ 2.5~cm) plastic scintillator,
were placed between the poles of the magnet to detect the $\beta$
particles emitted during the decay of $^{35}$K. Two identical {\em rf}
coils in a Helmholtz-like geometry were placed within the magnet and
the $\beta$ telescopes, with their field direction perpendicular to
both the direction of the beam and the static magnetic field. The
4~mm-thick, 22~mm-diameter disc-shaped KBr single crystal was mounted on
an insulated holder, between the pair of the {\em rf} coils and at a
45$^\circ$ angle with respect to the normal beam axis to minimize the
energy loss of the emitted $\beta$ particles. The spin-relaxation
time of the implanted $^{35}$K ions in the ionic crystal is much
longer than the decay lifetime~\cite{Matsuta92:K37rlx}.

For the initial asymmetry measurements to confirm nuclear polarization
of the ground state, the holding magnetic field of the dipole magnet
was switched ON and OFF every 60~s, at a maximum value while on equal
to 0.1~T, and the {\em rf} was always switched off~\cite{Anthony00:bnmr}.
The asymmetry ratio
\begin{equation}
R=\frac{[N(0^\circ)/N(180^\circ)]_{ON}}{[N(0^\circ)/N(180^\circ)]_{OFF}}=\frac{1+AP}{1-AP}
\label{eq:ratio}
\end{equation}
was then deduced by the counts detected in the $\beta$ scintillators,
$N(0^\circ)$ and $N(180^\circ)$, at directions 0$^\circ$ and
180$^\circ$, respective to the direction of the static magnetic field.
The asymmetry parameter in the $\beta$ decay of the $^{35}$K isotopes
is given as $A$ and $P$ is the polarization. The asymmetry ratio of
the pickup $^{35}$K products was determined near the center of the
outgoing momentum distribution of the fragment, $p_0$, namely at
$\frac{\Delta p}{p_0}=+0.5\%$. Two different momentum acceptance
settings, $0.5\%$ and $1\%$, were employed in combination with beam
angles of $1^\circ$ and $2^\circ$.

Once the asymmetry profile was established, the setting with the
maximum asymmetry was selected for the magnetic moment measurement, to
maximize the resonance effect during the frequency scans. The static
magnetic field was switched permanently ON at a value of 0.3012~T for
the NMR measurement, monitored by a Hall probe throughout the experiment.
A Hewlett-Packard HP 33120A Function Generator provided the {\em rf}
signal, which was amplified by an EIN 406L power amplifier. The coils
were configured as part of a RCL circuit, with a 50~$\Omega$ resistor
and a variable capacitor to match the output impedance of the {\em rf}
source with the input impedance of the coil, thus maximizing the
alternating magnetic field to the sample. The impedance of the
{\em rf}-coil was 59~$\mu$H. The strength of the {\em rf} signal was
monitored by measuring the voltage drop across the 50~$\Omega$ resistor
by means of an AC-DC voltage probe (Pomona 6106). The oscillating field
strength was maintained at $\approx 0.3$~mT. The frequency scans were
conducted using a frequency modulation (FM) of $\pm$10~kHz. The region
of frequency scans was limited between 520~kHz and 620~kHz, spanning
the full range around the previously measured $g$ factor of
$^{35}$K~\cite{Schaefer98:k35}.

Systematic asymmetries in the setup were checked before and after the
run with a standard $^{22}$Na source that was placed in the center of
the \bnmr apparatus, at the same position with the KBr crystal.


\section{Results}
\label{results}

\subsection{Asymmetry measurements}
\label{asymmetry}
\begin{figure}
\begin{center}
\includegraphics[
width=8cm
]{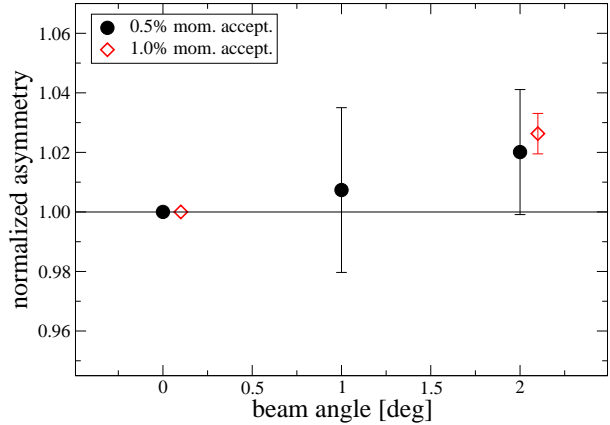}
\end{center}
\caption{(Color online) Measured asymmetry as a function of the outgoing momentum
of the $^{35}$K pickup products at $\Delta p/p=+0.5\%$. The open diamonds
are slightly displaced to the right for clarity.}
\label{fig:polres}
\end{figure}
The polarization, $P$, is directly related to the asymmetry ratio,
$R$, and the asymmetry parameter, $A$, as shown in
Eqn.~\ref{eq:ratio}. For a measurement of $P$, $R$ is established from
the data and $A$ can be deduced from the decay scheme. However, the
decay scheme of $^{35}$K (Fig.~\ref{fig:k35decay}) is inconclusive
about the asymmetry parameter. From the known properties of the decay,
only limits of $A$ can be imposed. Hence, we were unable to deduce the
absolute polarization of the $^{35}$K fragments. The asymmetry
measurements were completed for two different momentum acceptance
points, as described earlier, and the results are depicted in
Fig.~\ref{fig:polres}.

Normalization is required to correct for systematic asymmetries
existing in the setup, mainly due to the fringing magnetic field
affecting the photomultipliers. The photomultiplier tubes were
shielded with $\mu$ metal and soft iron shields, however, a small
asymmetry ($<1\%$) was noted in field ON/OFF measurements with
unpolarized radioactive sources. Measurements at 0$^\circ$ produce
nuclei with no polarization and can serve as normalization points for
all measurements.
\begin{figure}
\begin{center}
\includegraphics[
width=8cm
]{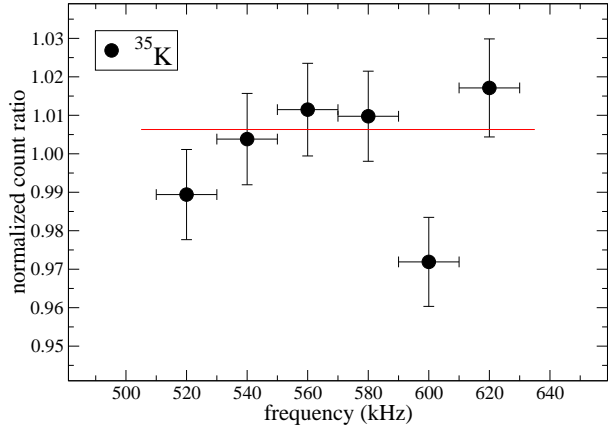}
\end{center}
\caption{(Color online) Results from the frequency scans to detect the resonance
of $^{35}$K fragments. The straight line is the linear fit to the offline
source data. The vertical error bars correspond to statistical errors,
while the horizontal ones correspond to the frequency modulation step.}
\label{fig:resonance}
\end{figure}

The data at 2$^\circ$ and $1\%$ momentum acceptance present an
asymmetry equal to $(2.8 \pm 0.8)\%$, produced in the charge-pickup
reaction for the $^{35}$K fragments. The result comes as a
confirmation of the previous observation of polarization in the case
of $^{37}$K nuclei~\cite{Groh03:k37} produced in a proton-pickup
reaction.

The observed asymmetry is sufficient to allow a measurement of the
ground-state magnetic moment in $^{35}$K using the $\beta$-NMR
technique. We note that the measured asymmetry includes the influence of the
$^{34}$Ar contamination in the secondary beam. $^{34}$Ar $\beta$
decay should show no asymmetry due to the $0^+$ ground state.
The decay properties ($J^\pi=0^+$, $t_{1/2}=844.5$~ms, $Q_{EC}=6061$~keV)
are similar to $^{35}$K. Thus, the observed asymmetry ratio is reduced.

There are two additional measurements at 0.5$\%$ momentum acceptance
and 1$^\circ$ and 2$^\circ$, respectively. The experimental error is
large for both measurements, crossing the no-asymmetry line and making
the results inconclusive. However, an increasing trend in the
asymmetry as the beam angle becomes larger is apparent, while the
value at 0.5$\%$ and 2$^\circ$ seems to agree within statistical accuracy,
with the measured asymmetry for 1$\%$ and 2$^\circ$.

\subsection{Magnetic moment of $^{35}$K }
\label{moment}

The magnetic moment measurement was carried out with the same settings
in the primary beam and the A1900 fragment separator that produced the
asymmetry of 2.8$\%$ in the prior asymmetry measurement, i.e. 2$^\circ$
beam angle and $1\%$ momentum acceptance. The resonance was found at a
frequency $\nu_1=600 \pm 10$~kHz, after sweeping the region 520--620~kHz
(Fig.~\ref{fig:resonance}). A statistical significance of 3$\sigma$ away
from the reference baseline was found for the resonance. The baseline was
determined independently during the $^{22}$Na source runs. The overall
statistical certainty of the present measurement reaches a level of 99.7$\%$.

From the frequency of the resonance, the corresponding $g$ factor was
deduced as $0.261(5)$, where the quoted error originates from the
frequency modulation width. The magnetic moment can be further
extracted as $\mu = g J$, with $J=3/2$ being the spin of the $^{35}$K
ground state~\cite{Ewan80:K35}. The final result is $|\mu$($^{35}$K)$|$=0.392(7) $\mu_N$.


\section{Discussion}
\label{discussion}

\begin{figure}
\begin{center}
\includegraphics[
width=8cm
]{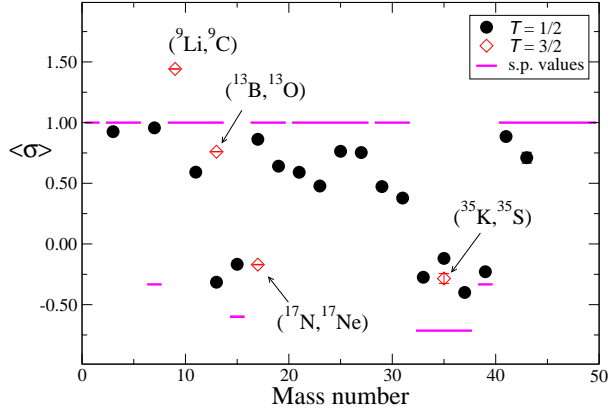}
\end{center}
\caption{(Color online) Systematics of isoscalar magnetic moments for both $T=1/2$
and $T=3/2$ mirror nuclei. The labels and arrows refer to the $T=3/2$
data known to date.}
\label{fig:systematics}
\end{figure}
The deduced spin-polarization of $^{37}$K nuclei at intermediate
energies was measured at 8.5$\%$~\cite{Groh03:k37} and with a positive
sign, confirming the fact that the picked-up proton from the
production target is preferentially located at Fermi momentum and with
its momentum aligned with the axis of the incident projectile. The
asymmetry parameter $A$ in Eqn.~\ref{eq:ratio} is known for $^{37}$K,
enabling the translation of the measured asymmetry ratio $R$ into
polarization. In $^{35}$K nuclei the asymmetry parameter is not known,
and the asymmetry is obscured in part by the presence of $^{34}$Ar
contamination. It is noted that significant $\beta$ asymmetry has
been observed for $^{35}$K fragments, suggesting that spin polarization
is maintained even with evaporation of two neutrons from proton
pickup products.

The measured value of the ground state $g$ factor, 0.261(5) falls in
range with the $g$ factor, $g=0.24(2)$, previously measured with a
much larger error. The Schmidt prediction for the $g$ factor of a
single proton in 1$d_{3/2}$ is $g=0.083$, acting as an
extreme limit for the measured $g$ factor. Previous theoretical
predictions for $^{35}$K include the shell-model prediction using the
USD interaction, $g_{USD}=0.122$, and an ``effective'' shell-model
calculation $g_{USD}^{eff}=0.243$ as quoted in Ref.~\cite{Schaefer98:k35}.
The present measurement is in good agreement with the prediction
employing the effective USD interaction. A configuration mixing which
was suspected to exist for the $T=3/2$ seems to be confirmed~\cite{Sherr75:calc}.
A simple $Z/A$ collective prediction, $g_{coll}=0.542$, doubles the
experimental value, thus promoting the dominant single-particle nature
in the ground state wavefunction of $^{35}$K. The sign of $g$ can not
be determined directly from the current measurement. However, we assume
it to be positive based on theoretical considerations for a single
proton in the 1$d_{3/2}$ level.

The existing data for the $^{35}$S nucleus may be combined to the
present result to extract the Pauli spin expectation value, $\langle
\sigma \rangle$, for the mirror pair at $A=35$. Using
Eqn.~\ref{eq:spin} a value $\langle \sigma \rangle=-0.284(40)$ is
calculated. Fig.~\ref{fig:systematics} depicts the systematics for all
available $T=1/2$ and $T=3/2$ data for $\langle \sigma\rangle$ as a
function of mass number. The $^{35}$K-$^{35}$S value is the heaviest
$T=3/2$ mirror pair known to date and agrees with the systematics
of $T=1/2$ nuclei. The deviation of $\langle \sigma\rangle$ away from
the Schmidt value is attributed mainly to core polarization effects
reflected in the magnetic moments of both $^{35}$K and $^{35}$S.

An interesting approach to the relationship between gyromagnetic
ratios and the strengths of the $\beta$-decay transitions of mirror
nuclei in the region $3 \le A \le 43$ has been followed by Buck {\em
et al.}~\cite{Buck83:older,Buck01:newer}. At the extreme case of {\em
only} odd nucleons being active, the spin dependence of the proton and
neutron magnetic moments can be eliminated and a linear relationship
of the form $\gamma_p$=$\alpha \gamma_n + \beta J$ between mirror
partners magnetic moments can be constructed. The coefficients
$\alpha$ and $\beta$, besides being the slope and the intercept of the
straight line, are directly linked to important structure quantities:
$\alpha=(G_p-g_p)/(G_n-g_n)$ and $\beta=g_p-\alpha g_n$.
Here, $g_p=1$, $g_n=0$ are the orbital magnetic moments for proton and
neutron, while $G_p=5.586$ and $G_n=-3.826$ are the corresponding spin
moments. All magnetic moments are in units of the nuclear
magneton. $\gamma_{p,n}$ is the magnetic moment $\mu_{p,n}$ divided by
the spin, $J$. Linear fits of available data in mirror nuclei may be
worked out to obtain the estimates of $\alpha$ and $\beta$.

In the original approach by Buck and Perez, only data for the case of
$T=1/2$ were included, since there were only two mirror pair moments
known in $T=3/2$ nuclei. Data for $T=3/2$ nuclei existed only for mass
$A=9$ and mass $A=13$, i.e. the mirror pairs $^{9}$Li-$^9$C and
$^{13}$Be-$^{13}$O, respectively.

\begin{figure}
\begin{center}
\includegraphics[
width=8cm
]{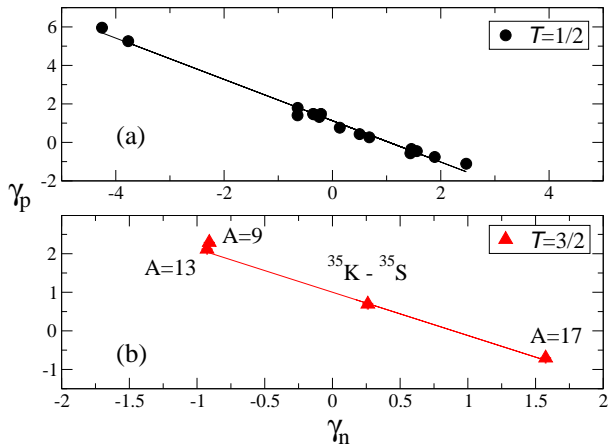}
\end{center}
\caption{(Color online) Plot of $\gamma_p$ vs. $\gamma_n$ according to~Ref.~\cite{Buck01:newer}
for all the available-to-date experimental magnetic moments data for
mirror nuclei. In (b), the linear fit does not include the
$^{9}$Li-$^{9}$C pair. The value for the mirror pair $^{35}$K-$^{35}$S
is the heaviest known to date. Please, note that the scale in the two
plots is different and the error bars are smaller than the size of the
plotted points.}
\label{fig:buck}
\end{figure}

With addition of the present work and a recent measurement in
$\mu$($^{17}$Ne)~\cite{Geithner05:ne_g} which completed the mirror
pair at $A=17$, and a remeasurement of the magnetic moment in
$^{9}$Li~\cite{Borremans05:Li9}, the Buck-Perez analysis was applied
for $T=3/2$ nuclei (Fig.~\ref{fig:buck}). Sch{\" a}fer {\em et al.} had expressed doubts on
the similarity of the behavior of $T=3/2$ mirror nuclei with the
systematics of $T=1/2$ data, based on their less precise measurement
for the $^{35}$K magnetic moment. It is therefore important to examine
the validity of the linear fits in $T=3/2$ nuclei and compare the
results to the slope and intercept from the $T=1/2$ systematics.

A linear fit was performed for all available $T=3/2$ data. The result
is shown in the third row of Table~\ref{tab:linear_fits}, together with
the theoretical ``bare'' nucleon values for $\alpha$ and $\beta$ (first
row) and the values obtained by Buck and Perez for the case of $T=1/2$
data (second row).
In recent studies~\cite{Huhta98:C9,Utsuno04:C9}, there have been strong
suggestions of an existing anomaly in $A=9$, which is responsible for
the large deviation of the corresponding $\langle \sigma \rangle$ value,
as can be seen in Fig.~\ref{fig:systematics}. The existence of such an
anomaly motivated a second regression analysis on the available $T=3/2$
data, excluding the $A=9$ point. The resulting values for the coefficients
$\alpha$ and $\beta$ in this case are depicted in the last row of
Table~\ref{tab:linear_fits}.
\begin{table}
\caption{Summary of the results of linear fits for the slope and
intercept according to Buck and Perez analysis.}
\label{tab:linear_fits}
\begin{ruledtabular}
\begin{tabular}{lcc}
isospin ($T$) & slope ($\alpha$) & intercept ($\beta$)\\
\hline
\vspace{-2mm}\\
1/2 (th.)& -1.199 & 1.0 \\
1/2 (exp.)~\footnote{Data taken from Ref.~\cite{Buck01:newer}} & -1.148 $\pm$ 0.010 & 1.052 $\pm$ 0.016 \\
3/2 (exp.)~\footnote{The fit includes the $A=9$ data}					 & -1.176 $\pm$ 0.087 & 1.063 $\pm$ 0.090 \\
3/2 (exp.)~\footnote{The fit excludes the $A=9$ data}					 & -1.122 $\pm$ 0.096 & 1.001 $\pm$ 0.102 \\
\end{tabular}
\end{ruledtabular}
\end{table}

Both fits are obtained with a very good correlation coefficient ($R=0.996$),
even with a limited number of points included in the regression. The available
$T=3/2$ data exhibit a linear trend, contradicting the argument by Sch{\" a}fer
{\em et al.} about their behavior. Also, the extracted values of $\alpha$ and
$\beta$ agree with the corresponding values in the case of $T=1/2$ nuclei.
In addition, the error bars of the slope and the intercept overlap with
the ``bare'' nucleon values. This is the first time a similarity in the
regression results for the $T=1/2$ and $T=3/2$ is observed.

The statistical uncertainty in the $T=3/2$ fits, however, is not sufficient
to allow any judgment about the role of the $A=9$ anomaly, as the extracted
values for $\alpha$ and $\beta$ coefficients have overlapping statistical errors.

Buck and Perez in their work~\cite{Buck83:older} deduced the relations
assuming that the contributions of even number particles in the
nuclei account for tiny (ideally zero) contributions. If this is true,
then no variation is expected between the $T=1/2$ and $T=3/2$ mirror
pairs, since their only first-order structural difference is a pair of
protons or neutrons. At this mass region, protons and neutrons occupy
similar orbitals, which is still true for $T=3/2$ nuclei. To first
order, the $p$--$n$ interaction is not expected to change dramatically
as isospin changes. At a higher order, $p$--$n$ interactions may play
a role, a consideration that is not adopted in the Buck-Perez
analysis. A deviation from the bare values would signify mainly the
effects of meson currents in the nuclei, with an overall effect of
screening the actual free nucleon current contribution in forming the
bare result for the total magnetic moment of the nucleus. Therefore,
it is important to improve on the statistical uncertainties for the
slope and intercept extracted from an analysis of $T=3/2$ mirror moment
data. This will require more measurements in the future.


\section{Conclusions}
\label{conclusions}

A proton-pickup ($p$,$2n$) reaction was employed to study the spin asymmetry
in polarized $^{35}$K nuclei produced from a $^{36}$Ar beam at intermediate
beam energies. The asymmetry measurements were carried out at various
momentum acceptance settings in the A1900 fragment separator at NSCL and
various beam angles. The asymmetry result obtained at 2$^\circ$ beam
angle and 1$\%$ momentum acceptance was $(2.8 \pm 0.8)\%$, confirming
the production of polarization in proton-pickup reactions, even with the
concurrence of neutron evaporation.

At the maximum observed asymmetry setting the \bnmr technique was employed
to measure the ground-state magnetic moment of the neutron-deficient $^{35}$K.
The extracted value, $|\mu$($^{35}$K)$|$=0.392(7), agrees with the previously
known value, but is improved significantly in the precision.

The measured magnetic moment was combined with the mirror nucleus magnetic
moment, $\mu$($^{35}$S), to extract the value of the Pauli spin expectation
value, $\langle \sigma \rangle=-0.284(40)$. The result is the heaviest
$T=3/2$ mirror pair known to date and agrees well with the systematics of
$T=1/2$ nuclei. Linear fits of the existing magnetic moments of $T=3/2$ nuclei
according to the approach by Buck and Perez produced values for the slope
$\alpha$ and intercept $\beta$. For the case of all $T=3/2$ data,
$\alpha = -1.176(87)$ and $\beta=1.063(90)$. The extracted values agree well
with the ones from $T=1/2$ nuclei, and also overlap with the corresponding
``bare'' nucleon values. It is the first time such a behavior is observed
for $T=3/2$ nuclei. More magnetic moment measurements in $T_z=-3/2$ nuclei
are needed for that purpose.


\begin{acknowledgments}
\label{acknowledgments}

The authors are grateful to Dr. D.E. Groh and Dr. A.E. Stuchbery for
their useful remarks and discussions during the early stages of this work
and Dr. K. Minamisono for interesting discussions.
The authors would also like to thank the NSCL operations staff for
providing the primary and secondary beams for this experiment.
This work was supported in part by the National Science Foundation
Grants PHY-01-10253 and PHY-99-83810.
 
\end{acknowledgments}



\bibliography{k35tjm}
\label{bibliography}

\end{document}